\documentclass[11pt]{article}
\usepackage{moriond,epsfig}
\begin{document}
\vspace*{4cm}

\title{BEYOND HEAVY TOP LIMIT IN HIGGS BOSON PRODUCTION AT LHC}
\author{Alexey Pak, Mikhail Rogal, and Matthias Steinhauser}
\address{Institut f{\"u}r Theoretische Teilchenphysik (TTP), KIT Karlsruhe}
\maketitle\abstracts{
QCD corrections to inclusive Higgs boson production at the LHC are 
evaluated at next-to-next-to leading order. By performing asymptotic 
expansion of the cross section near the limit of infinitely heavy top quark 
we obtained a few first top mass-suppressed terms. The corrections to 
the hadronic cross sections are found to be small compared to the 
scale uncertainty, thus justifying the use of heavy top quark
approximation in many published results.}

\section{Introduction}

The Large Hadron Collider (LHC) at CERN is expected to provide insights on 
the mechanism of electroweak symmetry breaking, possibly by 
discovering the elusive Higgs boson. In the Standard Model, the dominant
process of the Higgs boson production is the gluon fusion,
$gg\to H$, mediated by a top quark loop.
Predictions of Higgs boson production in gluon fusion
both at the Tevatron and the LHC~\cite{deFlorian:2009hc,Anastasiou:2008tj}
include electroweak effects and results beyond the fixed-order
perturbation theory, but QCD corrections have the greatest numerical effect.
Since 1977, when the leading order (LO) calculation appeared~\cite{WEGR},
also next-to-leading order (NLO)~\cite{Dawson:1990zj,Spira:1995rr,Djouadi:1991tka},
and more recently next-to-next-to-leading order (NNLO)
~\cite{Harlander:2000mg,Harlander:2002wh,Ravindran:2003um,Anastasiou:2002yz}
QCD corrections have been evaluated.

While the NLO results are exact in the top quark and Higgs boson masses,
the NNLO results rely on the effective theory built in the limit of the
large top quark mass
(for a review see e.g. Ref.~\cite{Steinhauser:2002rq}).
At NLO, this approximation results in $< 2\%$ deviations from the exact
result for $M_H<2M_t$~\cite{Harlander:2003xy,Kramer:1996iq}.
NNLO effects of the finite top quark mass have been first indirectly
addressed in Ref.~\cite{Marzani:2008az}, where the asymptotics in the opposite
limit of large center-of-mass energy $\sqrt{\hat{s}}$ were considered.
Recently, two independent groups~\cite{Harlander:2009mq,Pak:2009dg,Harlander:2009bw}
performed an expansion of the inclusive Higgs production cross-section in $\rho = M_H^2/M_t^2$.
In this contribution we summarize those results and provide some details of
our calculation~\cite{Pak:2009dg}.

\section{Calculation of partonic cross-sections}

The QCD corrections to the cross-sections of partons are:
\begin{eqnarray}
  \hat{\sigma}_{ij\to H + X} &=& \hat{A}_{\rm LO} \left(
    \Delta_{ij}^{(0)} + \frac{\alpha_s}{\pi}~ \Delta_{ij}^{(1)}  
    + \left(\frac{\alpha_s}{\pi}\right)^2 \Delta_{ij}^{(2)} + \ldots
  \right)
  \,,~~~
  \hat{A}_{\rm LO} = \frac{G_F~\alpha_s^2}{288\sqrt{2}\pi} 
  f_0(\rho,0).
  \label{eq::hatsigma}
\end{eqnarray}
Here $ij$ denote one of the possible initial states:
$gg$, $qg$, $q\bar{q}$, $qq$, or $qq^\prime$, and $q$ and $q^\prime$
stand for (different) massless quark flavours.
At the leading order, the only non-zero contribution is $\Delta_{gg}^{(0)} = \delta(1-x)$,
and the function $f_0(\rho,0)$~\cite{Pak:2009bx}
describes the mass dependence. We focus on the $x$- and $\rho$-dependence of
$\Delta^{(1)}_{ij}$ and $\Delta^{(2)}_{ij}$. As is common in the literature,
by ``infinite top quark mass approximation'' we assume that $\Delta_{ij}^{(k)}$ 
are evaluated for $M_t\to \infty$, but $\hat{A}_{\rm LO}$ is exact in $M_t$.
\begin{figure}[t]
  \centering
  \includegraphics[width=0.25\linewidth]{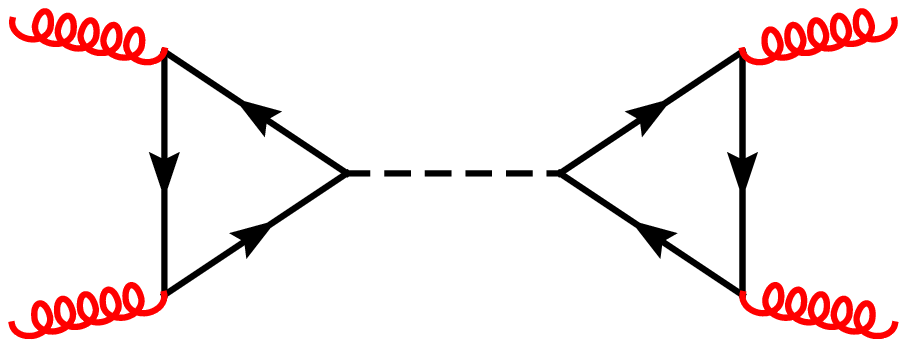}\hfill
  \includegraphics[width=0.25\linewidth]{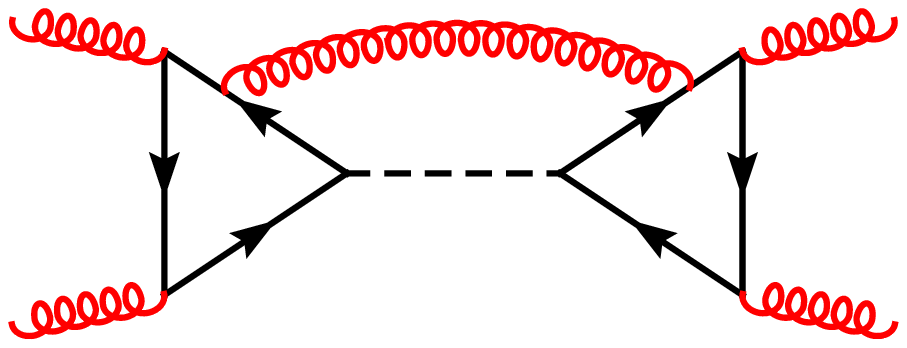}\hfill
  \includegraphics[width=0.25\linewidth]{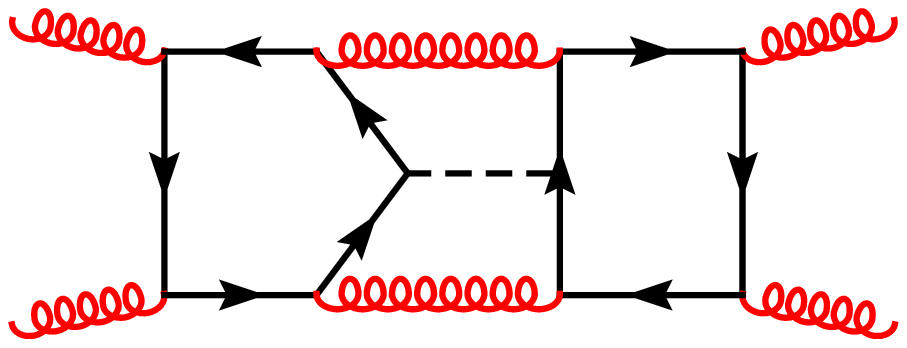}
  \caption[]{\label{fig::diag}Sample forward scattering diagrams
    whose cuts correspond to the LO, NLO and NNLO corrections to
    $gg\to H$. Dashed, curly and solid lines represent Higgs bosons,
    gluons and top quarks, respectively.}
\end{figure}

To account for the real and virtual corrections we employ the optical
theorem and compute imaginary parts of the four-point forward-scattering
amplitudes such as in Fig.~\ref{fig::diag}.
After the asymptotic expansion in the limit $M_t^2\gg\hat{s},M_H^2$
the loop integrals factorize. The most non-trivial cases are two-loop 
four-point functions dependent on both $\hat{s}$ and $M_H$.
Reducing them with IBP's~\cite{Laporta:2001dd} we obtain around 30 
master integrals. The latter are available~\cite{Anastasiou:2002yz},
however, we re-computed them with the combination of soft expansion and
differential equation methods. Finally, we add renormalization terms and
obtain a few first terms in the expansion of $\Delta^{(k)}_{ij}$ in 
powers of $\rho$, where coefficients are functions of $x$.

\section{NLO and NNLO results}

\begin{figure}[b]
  \begin{tabular}{ccc}
    \includegraphics[width=0.32\linewidth]{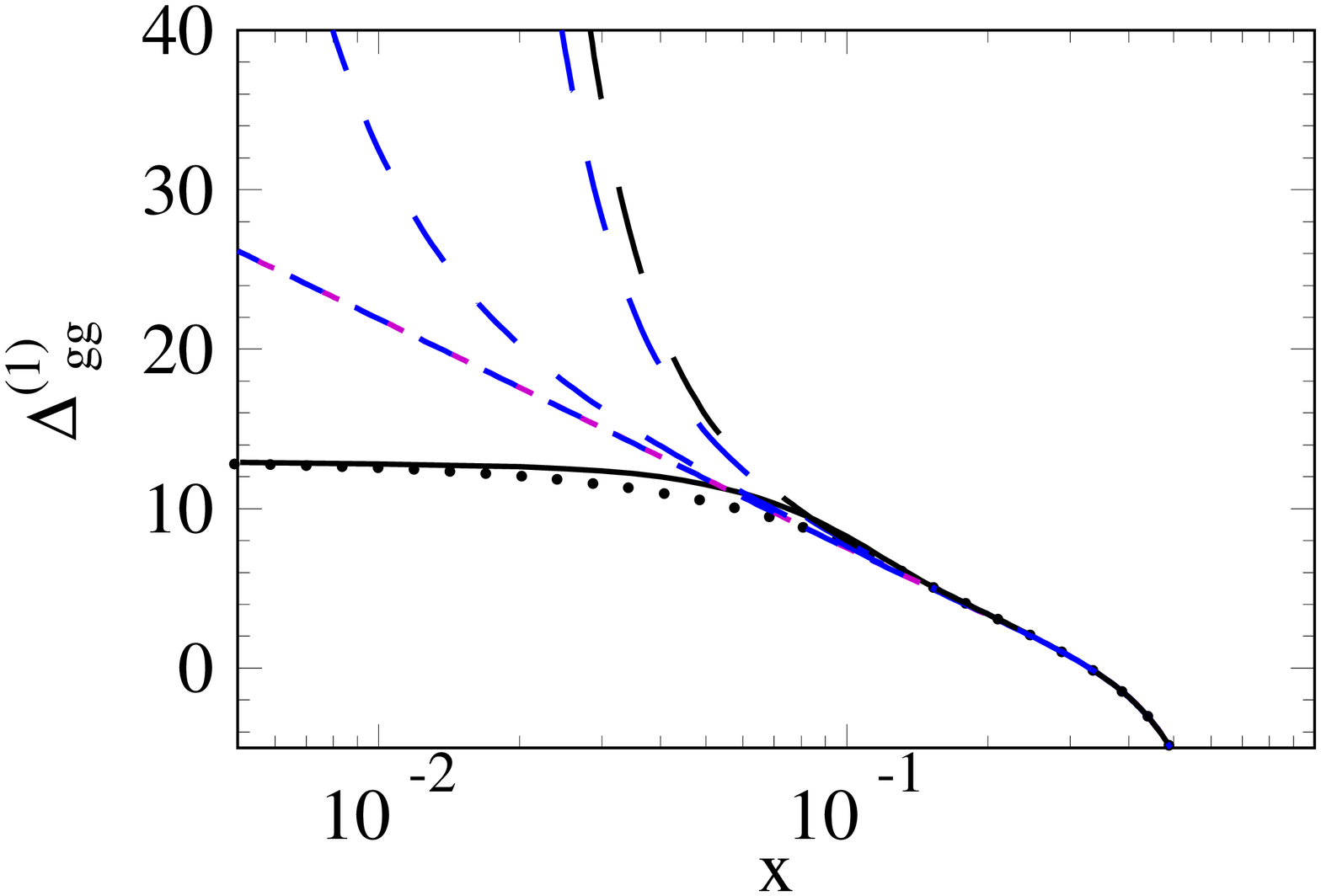}
    \put(-35,70){(a)} &
    \includegraphics[width=0.32\linewidth]{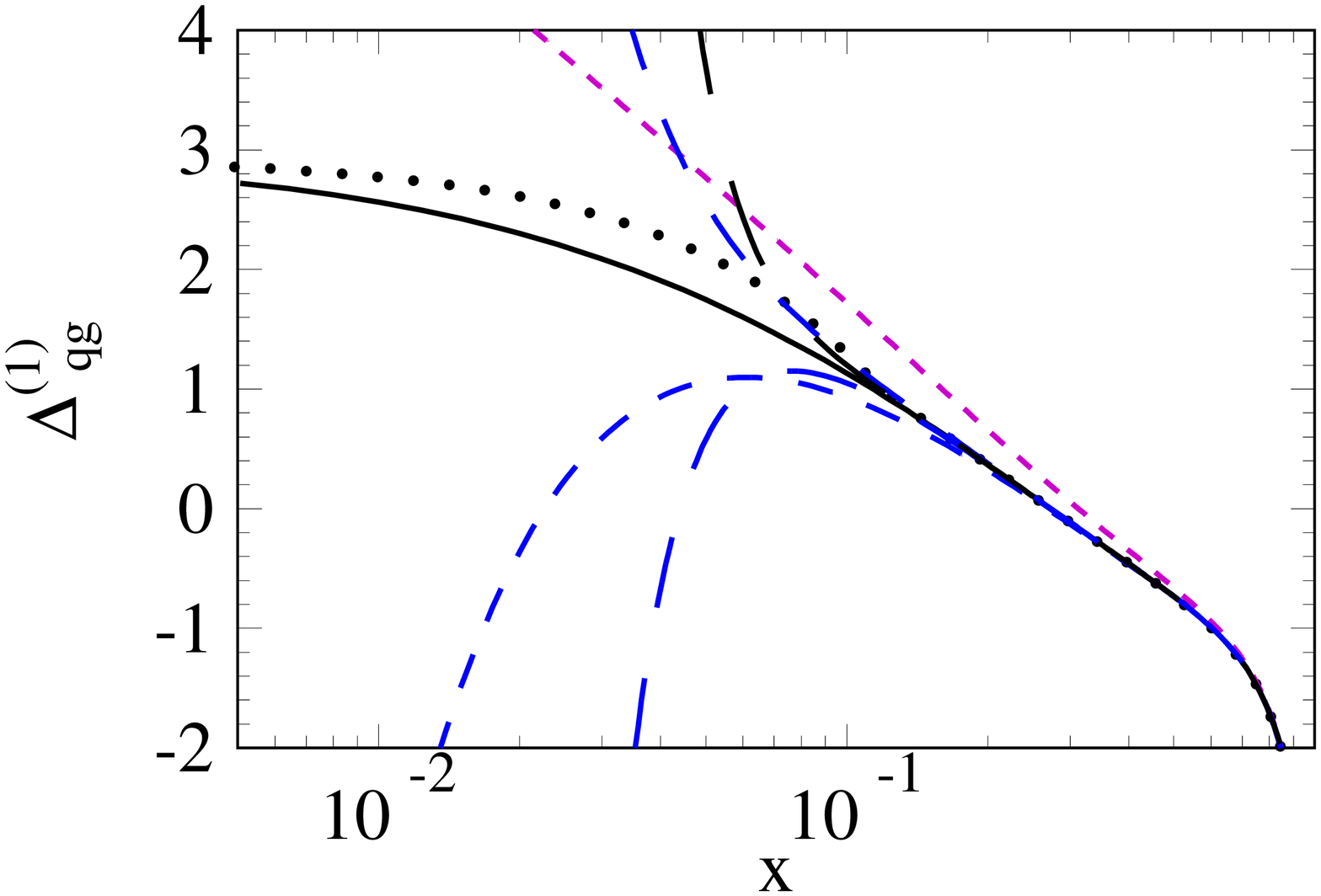}
    \put(-35,70){(b)} &
    \includegraphics[width=0.32\linewidth]{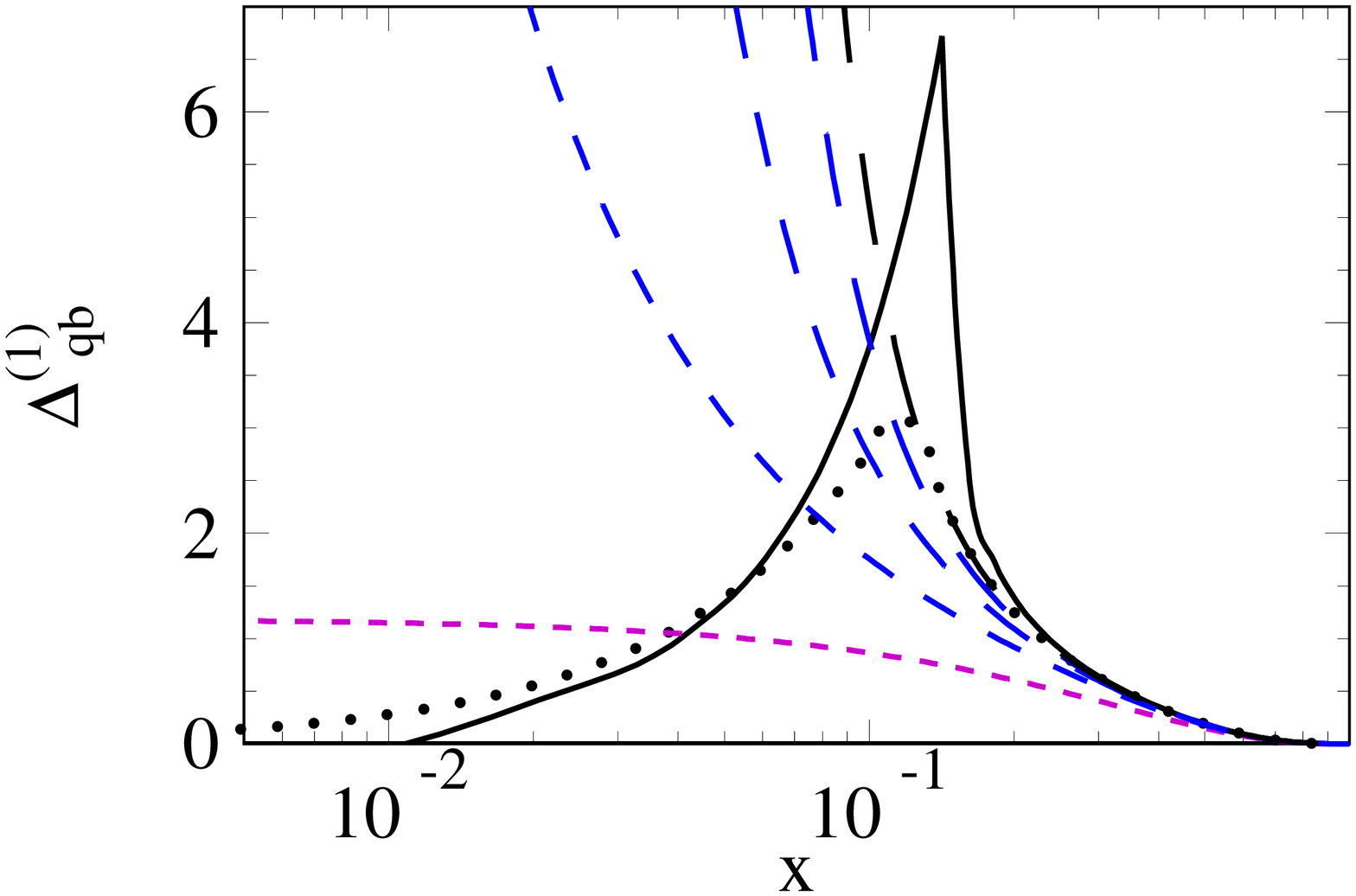}
    \put(-35,70){(c)}
  \end{tabular}
  \caption[]{\label{fig::NLOpart}NLO partonic cross sections for the
    (a) $gg$, (b) $qg$ and (c) $q\bar{q}$ channel as functions of $x$ for
    $M_H=130$~GeV. The expansion
    in $\rho\to0$ (dashed lines) is compared with the exact result (solid
    lines). Lines with longer dashes include higher order terms in $\rho$.
    The interpolation (see text) is shown as a dotted line.}
\end{figure}

In Fig.~\ref{fig::NLOpart} we compare the $x$-dependence of the exact NLO
results~\cite{Dawson:1990zj,Spira:1995rr,Djouadi:1991tka}
(evaluated for $M_H=130$~GeV and $M_t=173.1$~GeV) to the $\mathcal{O}(\rho^n)$
approximations for successive $n$.
The leading term in $\rho$ is smooth and demonstrates a reasonably
good agreement with the exact curve for $x\to 1$. However, the higher order
terms in $\rho$ introduce divergences at $x\to 0$ which are the
most obvious for the $q\bar{q}$ channel. This signifies the breakdown
of the assumption that $M_t^2\gg \hat{s}$ for large $\hat{s}$.
Note, however, the decent convergence above the threshold for the
top quark pair production
(in Fig.~\ref{fig::NLOpart}, $x_{\rm th} \approx 0.14$).
To recover the proper $x\to 0$ behaviour, we utilize $\hat{s}\to\infty$
asymptotics~\cite{Marzani:2008az,Harlander:2009bw}.
Interpolation between the $\mathcal{O}(\rho^n)$ result and the value at $x\to 0$
(dots in Fig.~\ref{fig::NLOpart}) agrees well with the exact curve
for the $gg$ channel. For the quark channels, the introduced error
in hadronic contributions does not exceed 50\%, which, if also true at NNLO,
translates to a shift less than the total scale uncertainty of the full NNLO cross-section.

The NNLO diagrams require considerably more effort.
Using the known virtual corrections~\cite{Harlander:2009bw,Pak:2009bx}
we were able to evaluate three terms in the expansion of $\Delta^{(2)}_{gg}$
and four terms in the other channels. Our analytic results are in full
agreement with the $M_t\to \infty$ results~\cite{Anastasiou:2002yz} and 
the mass corrections expanded in $(1-x)$ (soft expansion)
~\cite{Harlander:2009mq,Harlander:2009bw}.
\begin{figure}[t]
  \centering
  \begin{tabular}{ccc}
    \includegraphics[width=0.32\linewidth]{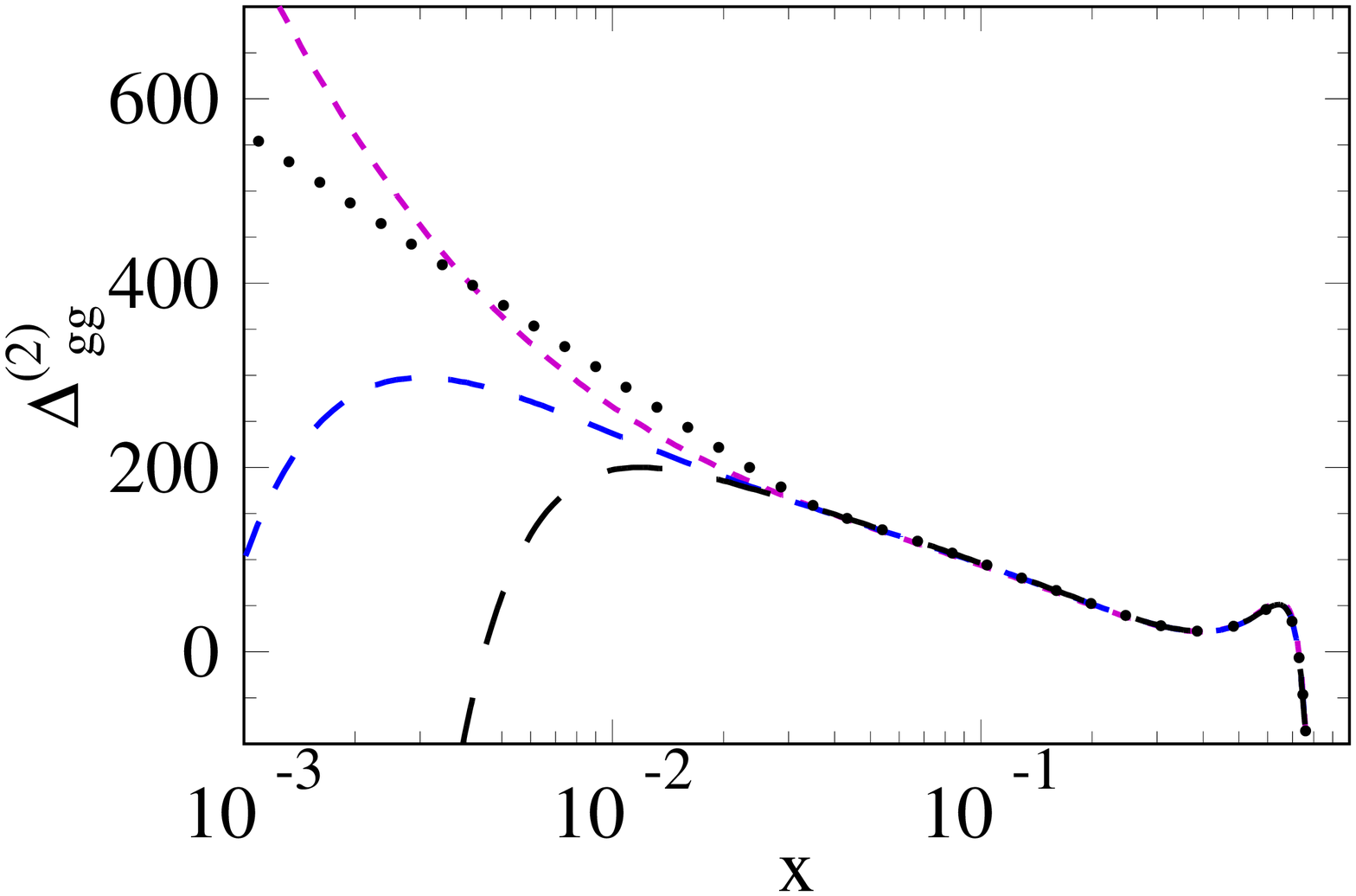}
    \put(-35,70){(a)} &
    \includegraphics[width=0.32\linewidth]{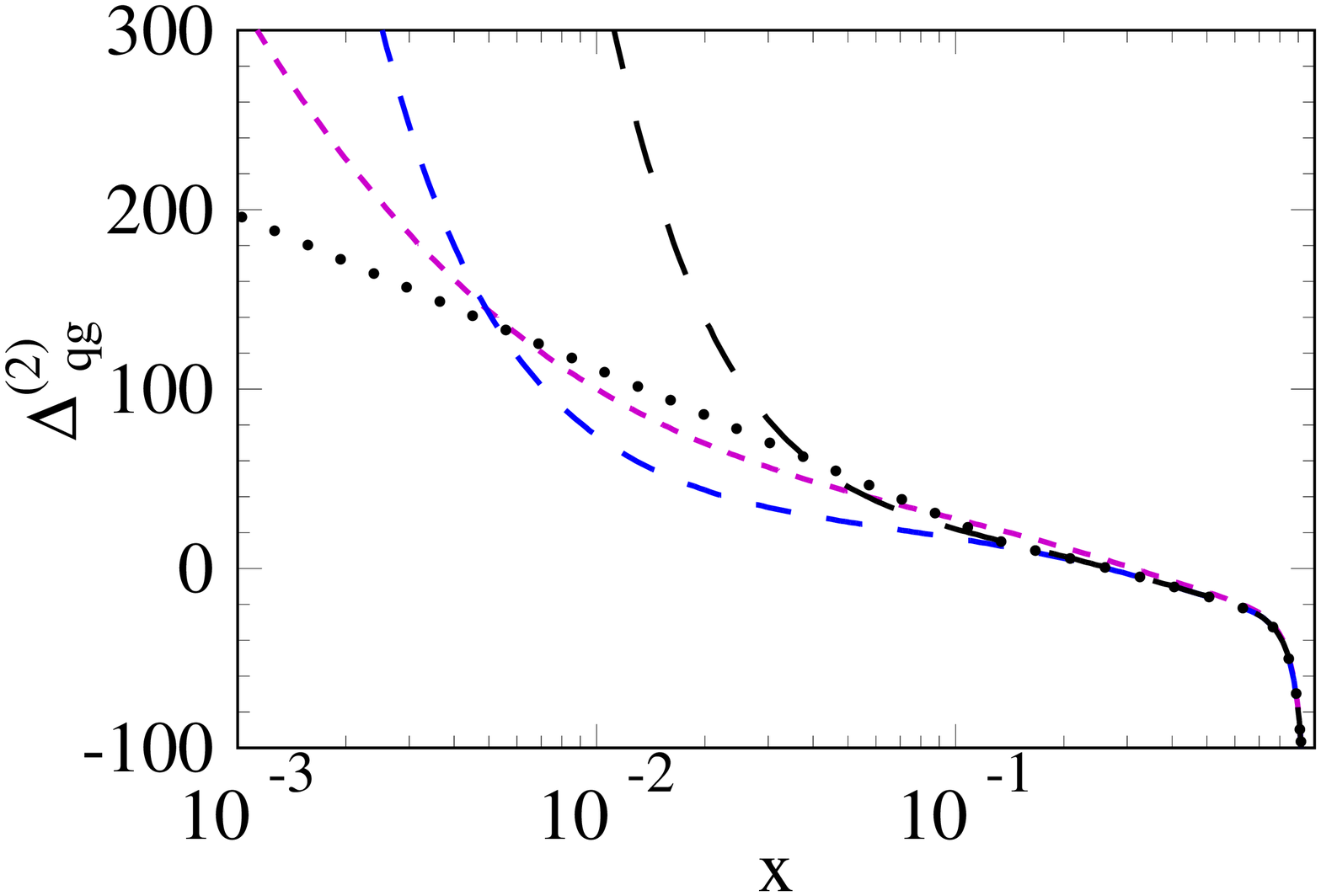}
    \put(-35,70){(b)} &
    \includegraphics[width=0.32\linewidth]{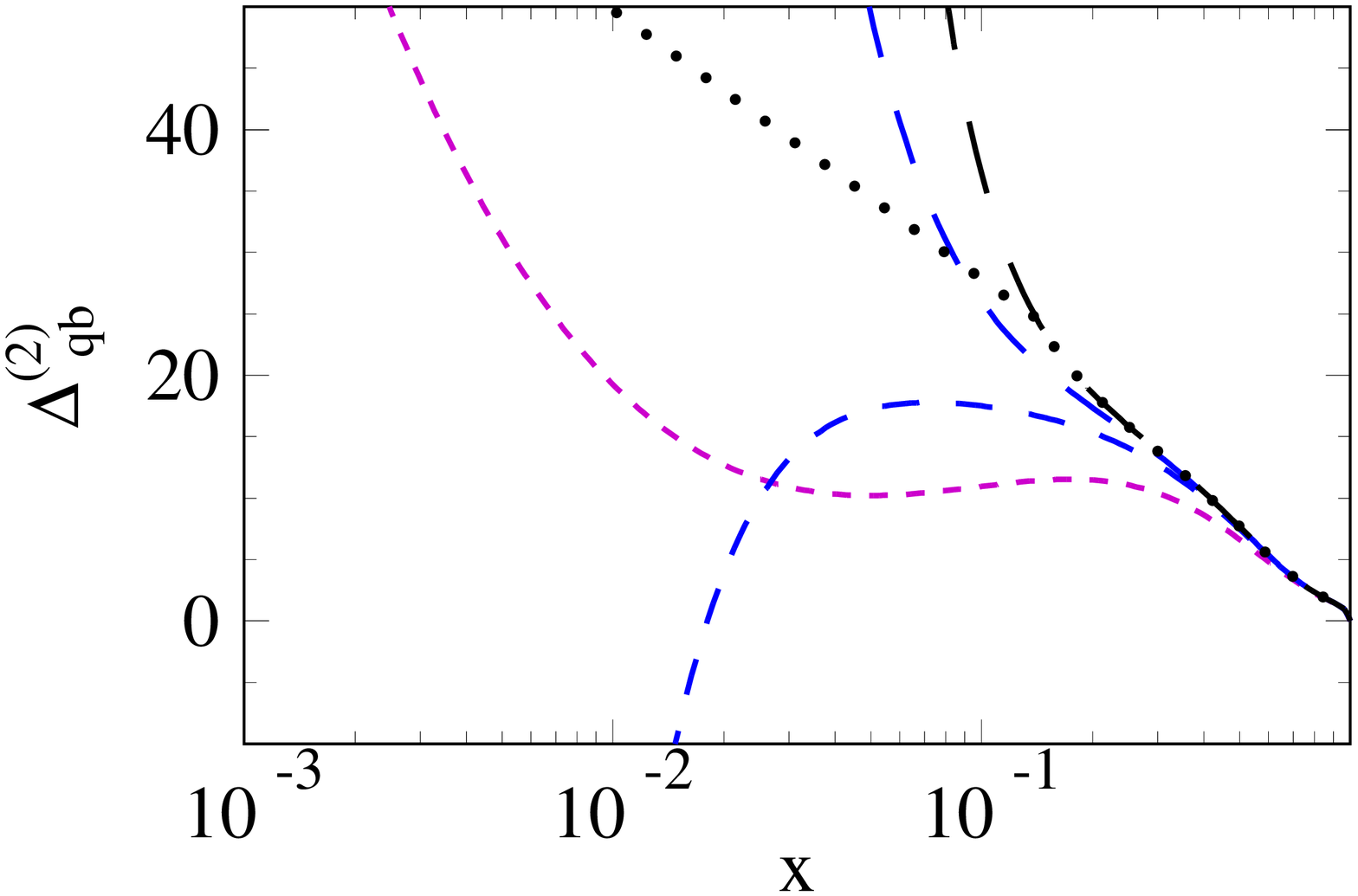}
    \put(-35,70){(c)}
  \end{tabular}
  \caption[]{\label{fig::NNLOpart}Partonic NNLO cross sections for
    the (a) $gg$, (b) $qg$, (c) $q\bar{q}$ channels
    for $M_H=130$~GeV.
    Lines with longer dashes include higher order terms in $\rho$.
    Dotted lines corresponds to interpolation.}
\end{figure}
In Fig.~\ref{fig::NNLOpart} we present $x$-dependence of the 
functions $\Delta^{(2)}_{gg}$, $\Delta^{(2)}_{qg}$, and $\Delta^{(2)}_{g\bar{q}}$,
with interpolations constructed similarly to the NLO case.

\section{Hadronic results}

\begin{figure}[b]
  \centering
  \begin{tabular}{ccc}
    \includegraphics[width=0.32\linewidth]{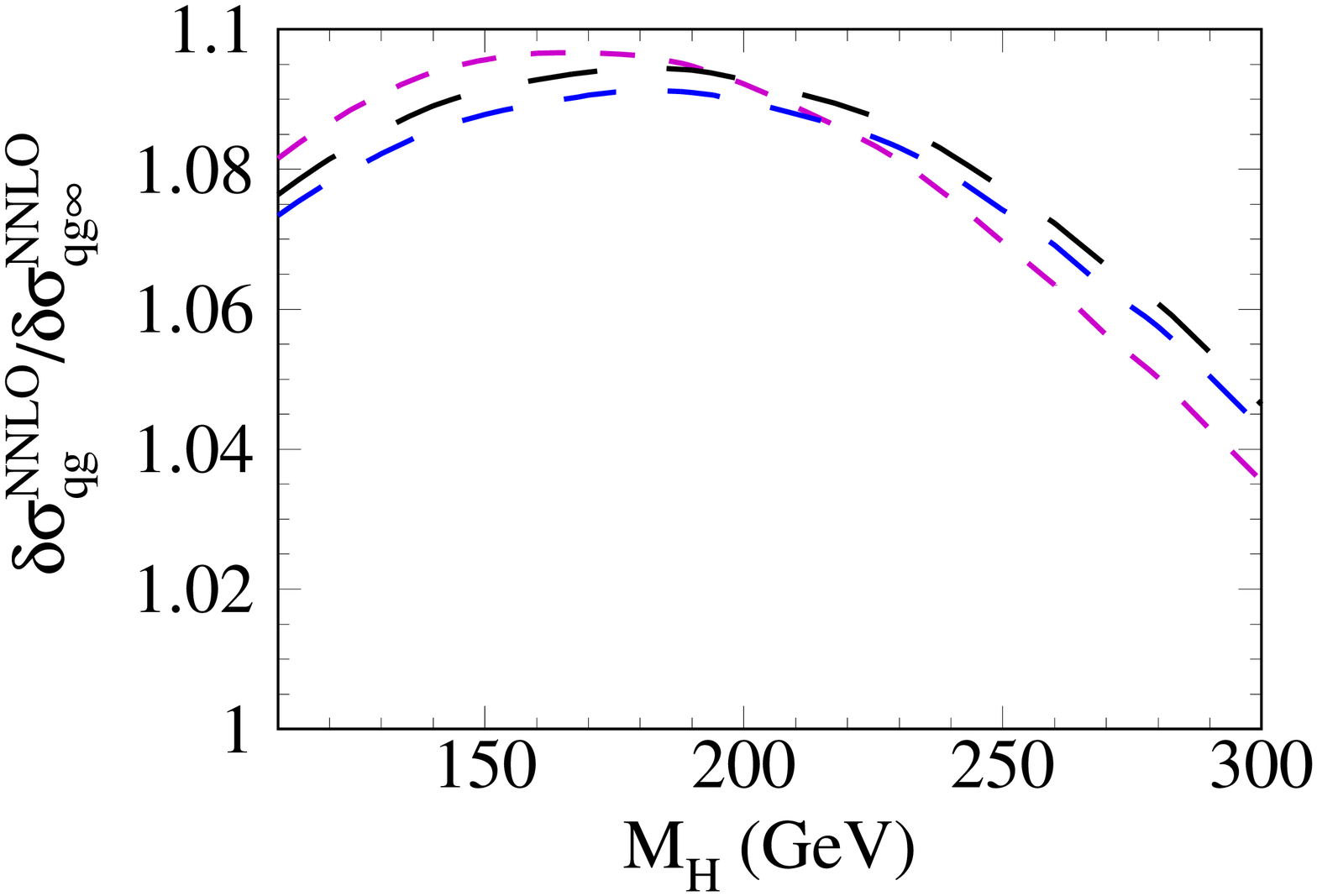}
    \put(-30,70){(a)} &
    \includegraphics[width=0.32\linewidth]{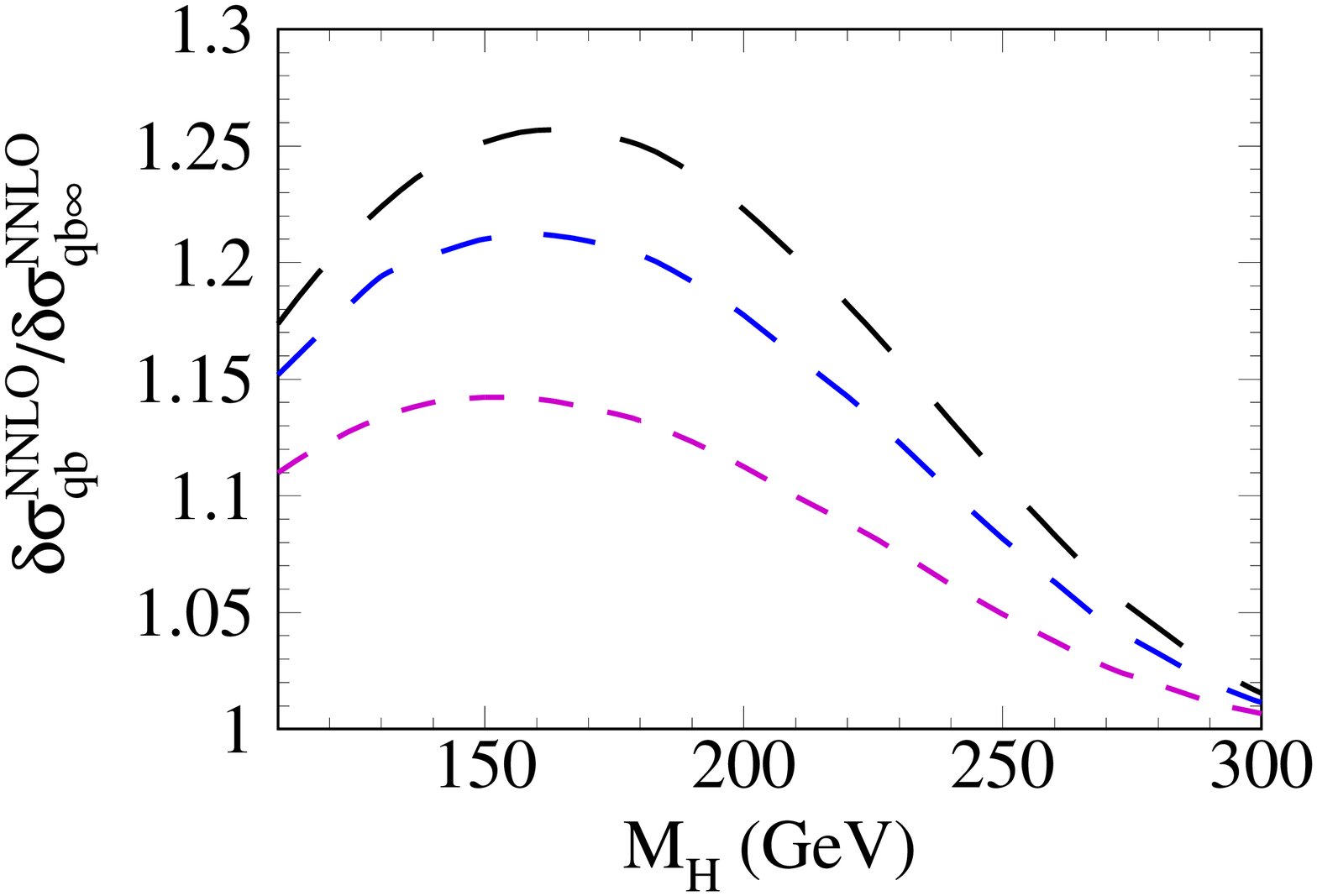}
    \put(-30,70){(b)} &
    \includegraphics[width=0.32\linewidth]{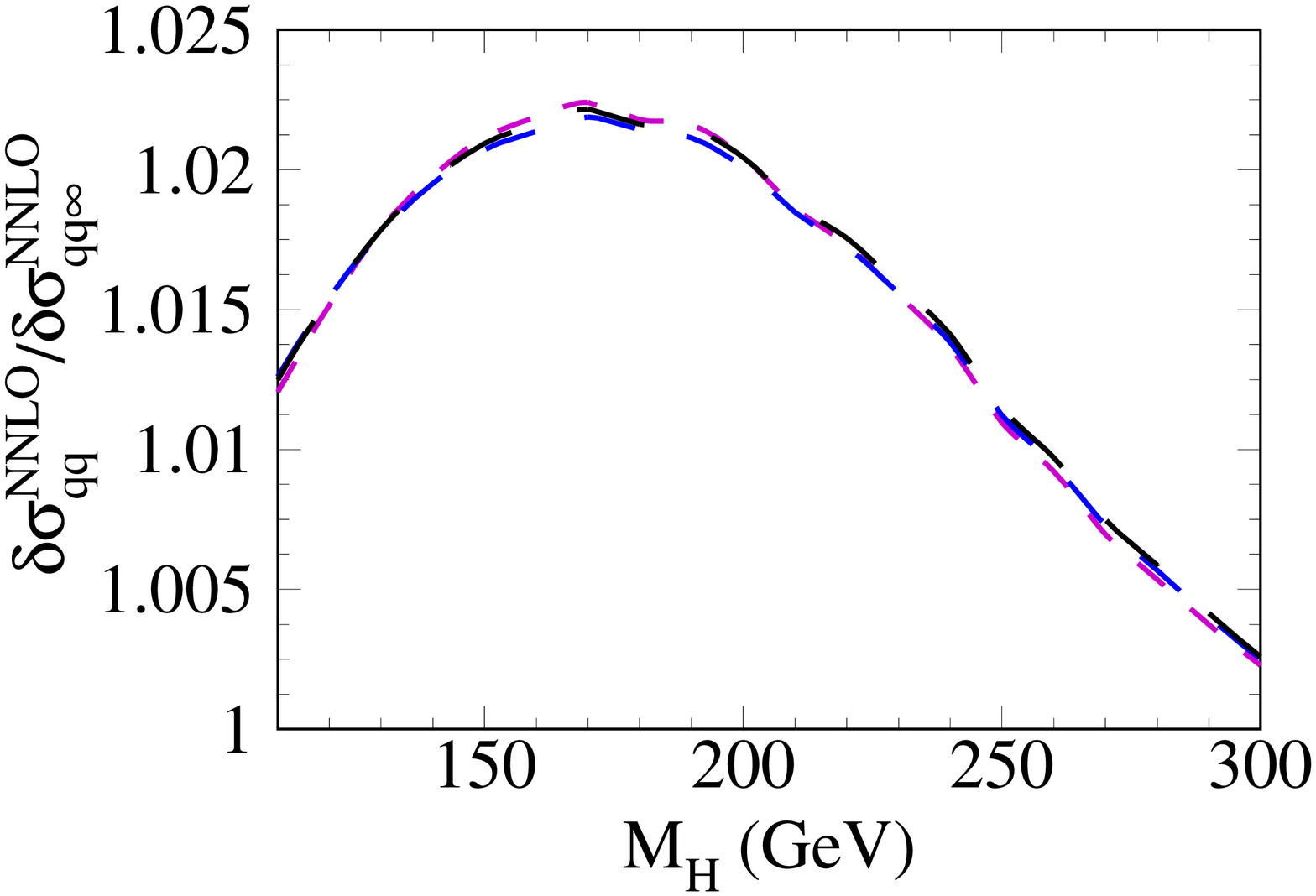}
    \put(-30,70){(c)} 
  \end{tabular}
  \caption[]{\label{fig::opt2_nnlo}NNLO contributions to hadronic cross section
    with higher orders in $1/M_t$ (from short to long dashes)
    normalized to the heave top quark mass result,
    (a) $qg$, (b) $q\bar{q}$, (c) $qq$. Channels $qq^\prime$ and $qq$
    are almost identical.}
\end{figure}

The hadronic cross sections are given by the convolution of 
$\hat{\sigma}_{ij\to H+X}$ with the corresponding
parton distribution functions (PDFs). We decompose it 
into LO, NLO, and NNLO contributions:
$\sigma_{pp^\prime\to H+X}(s) = \sigma^{\rm LO} + \delta\sigma^{\rm NLO} + \delta\sigma^{\rm NNLO}$.
In Fig.~\ref{fig::opt2_nnlo} we show the $M_H$-dependence of 
$\delta_{qg}^{(2)}$, $\delta_{q\bar{q}}^{(2)}$, and $\delta^{(2)}_{qq}$
normalized to the infinte top quark mass result, labeled with subscript $\infty$.
In all cases the power-suppressed terms lead to an increase of the cross
section between 4\% and 10\% for the quark-gluon and up to 25\% for the
quark-anti-quark channel in our range of Higgs boson masses.
For the $qq$ and $qq^\prime$ channels we observe rapid convergence
beyond $1/M_t^2$.

\begin{figure}[t]
  \centering
  \begin{tabular}{cc}
    \includegraphics[width=0.4\linewidth]{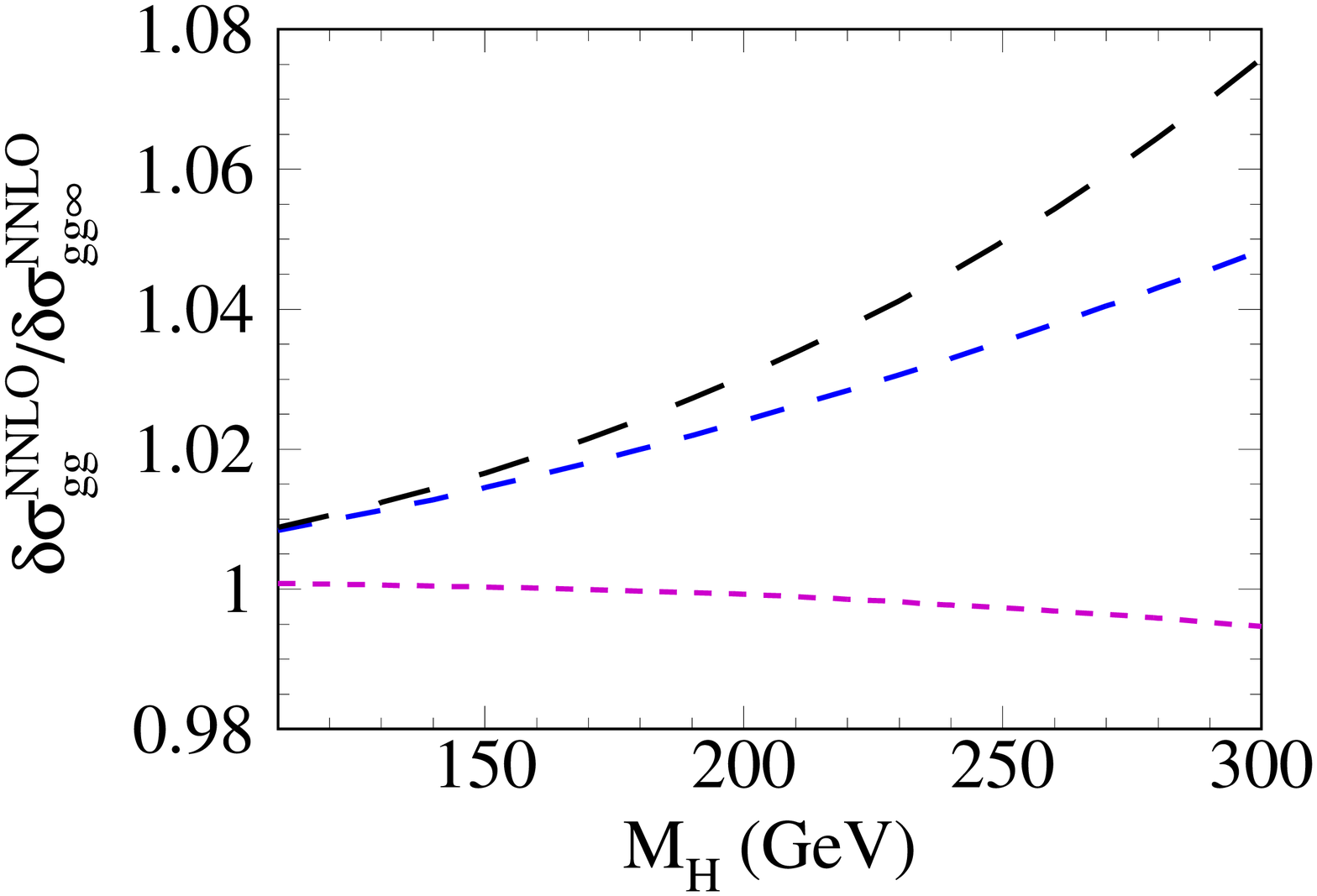}
    \put(-140,100){(a)} &
    \includegraphics[width=0.4\linewidth]{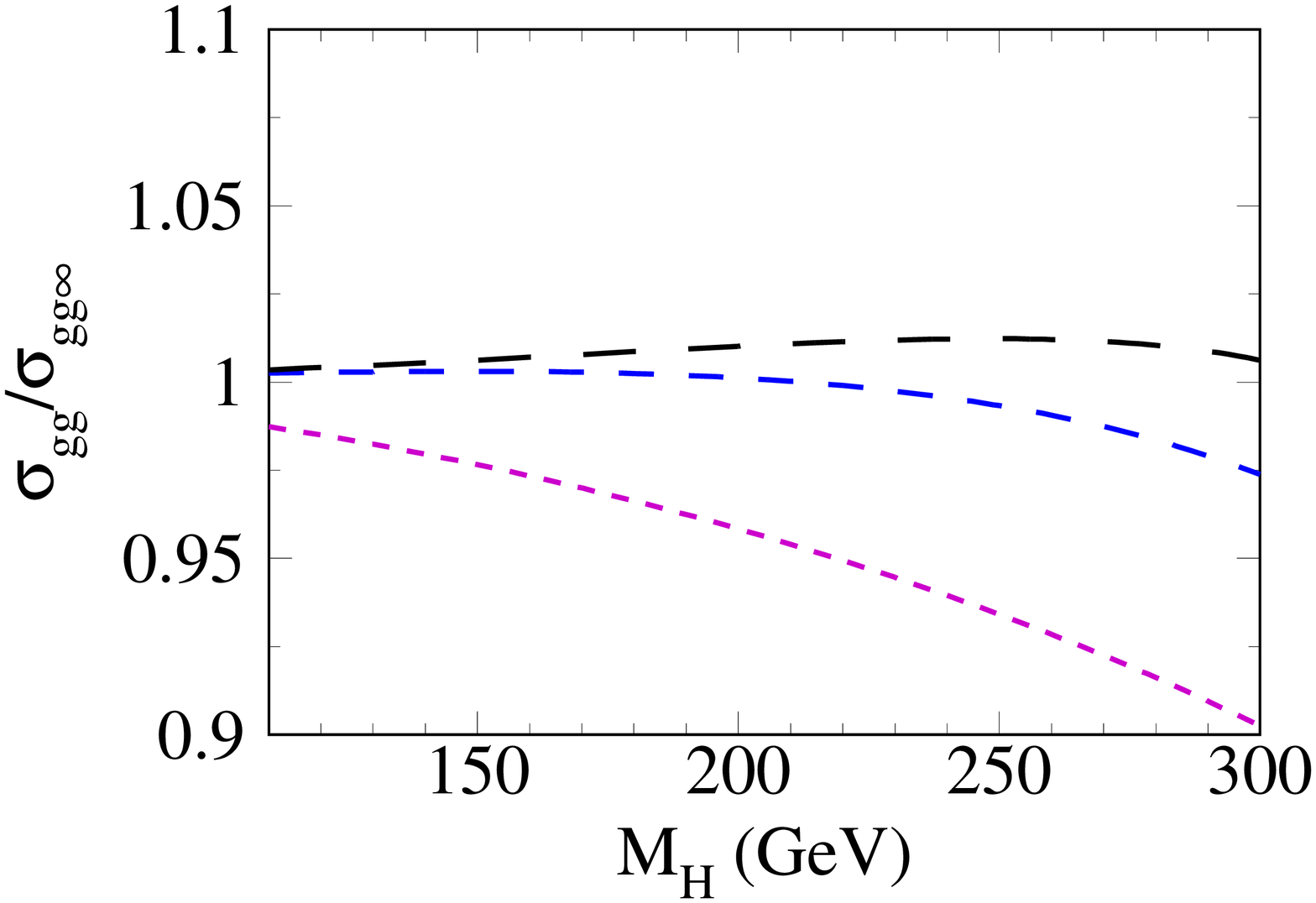}
    \put(-140,100){(b)}
  \end{tabular}
  \caption[]{\label{fig::opt1_nnlo}
    (a) Ratio of the NNLO hadronic cross section 
    ($gg$ contribution) including successive higher orders in $1/M_t$ normalized
    to the infinite top quark mass result. 
    (b) Prediction for the gluon-induced inclusive Higgs production
    cross section up to NNLO normalized to the heavy top limit.}
\end{figure}

NNLO corrections to the $gg$ channel are shown in 
Fig.~\ref{fig::opt1_nnlo}(a). Finally, in Fig.~\ref{fig::opt1_nnlo}(b) we 
present the gluon-induced cross-section including exact LO and NLO contributions.
Minor differences with the left panel of Fig.~7 in Ref.~\cite{Harlander:2009mq}
can be attributed to the different matching procedure. As one can see, the
effects of matching near $x=0$ and $M_t$-suppressed corrections nearly cancel
and the final deviation from the heavy top mass result is below 1\%
(when exact LO mass dependence is factored out).

\section{Conclusion}

We present the NNLO production cross section of the
Standard Model Higgs boson including the finite top quark mass effects.
To improve $x\to 0$ behaviour for the gluon-gluon channel we match our
results to the $\hat{s}\to\infty$ limit.
The numerical impact of the top quark mass suppressed terms is 
below 1\% and thus about a factor of ten smaller than the scale
variation uncertainty. Our calculation justifies the use of the heavy
top quark mass approximation in NNLO cross section calculations.
In addition, we independently confirm the analytic results in the heavy
top limit~\cite{Anastasiou:2002yz} and the soft expansion of
$M_t$-suppressed terms~\cite{Harlander:2009mq}.

{\itshape Acknowledgements.}
This work was supported by the DFG through the SFB/TR~9 ``Computational
Particle Physics''. M.R. was supported by the Helmholtz Alliance ``Physics at the Terascale''.

\end{document}